# An updated overview of radiomics-based artificial intelligence (AI) methods in breast cancer screening and diagnosis

Reza Elahi[1*], Mahdis Nazari[2]

[1]Department of Radiology, Zanjan University of Medical Sciences, Zanjan, Iran

[2]School of Medicine, Zanjan University of Medical Sciences, Zanjan, Iran

*Corresponding author's email: rezaelahi96@zums.ac.ir

## Abstract

Current imaging methods for diagnosing BC are associated with limited sensitivity and specificity and modest positive predictive power. The recent progress in image analysis using artificial intelligence (AI) has created great promise to improve breast cancer (BC) diagnosis and subtype differentiation. In this case, novel quantitative computational methods, such as radiomics, have been developed to improve the sensitivity and specificity of early BC diagnosis and classification. The potential of radiomics in improving the diagnostic efficacy of imaging studies has been shown in several studies. In this review article, we discuss the radiomics workflow and current hand-crafted radiomics methods in the diagnosis and classification of BC based on most recent studies on different imaging modalities, e.g. MRI, mammography, contrast-enhanced spectral mammography (CESM), ultrasound imaging, and digital breast tumosynthesis (DBT). We also discuss current challenges and potential strategies to improve the specificity and sensitivity of radiomics in breast cancer to help achieve a higher level of BC classification and diagnosis in the clinical setting. The growing field of AI incorporation with imaging information has opened a great opportunity to provide a higher level of care for BC patients.

# Introduction

The increasing number of breast cancer (BC) in recent years has made it the most common malignancy of women, with more than two million annual diagnoses worldwide (1). Despite all the efforts to optimize the screening and diagnosis of BC, its survival rate is 50-60%, and its mortality is assumed to increase to 43% by 2030 (2, 3). Therefore, improving screening and diagnostic methods to achieve an early diagnosis of BC is mandatory to improve the treatment outcome and survival. Based on the American College of Radiology (ACR) and Society of Breast Imaging (SBI) recommendations, all women must be assessed regarding BC by age 30. The annual mammographic screening started in women older than 40 can reduce mortality by up to 40%. The current guidelines do not consider an upper age limit for BC screening (4). Despite the considerable reduction of BC mortality via BC screening, late-stage diagnosis, the low specificity of current screening guidelines, and unnecessary biopsy sampling are still the main challenges (5).

Current guidelines for BC diagnosis are based on clinical evaluation of the patient and qualitative radiologic imaging using mammography, magnetic resonance imaging (MRI), ultrasound imaging, and digital breast tomosynthesis (DBT). These methods have multiple limitations, such as dependency on the skill and experience of the radiologist, suboptimal positive predictive value, and suboptimal sensitivity of the current imaging methods (approximately 70%) (6, 7).

To obtain a definitive diagnosis of BC, histopathologic evaluation of the biopsy sample, as the gold standard method of diagnosis, is required. Biopsy specimen obtaining is an invasive procedure, and its efficacy can be limited by the heterogeneity of BC, sampling errors, problematic sampling in small and necrotic tumors, beauty problems, and suboptimal accuracy of biopsy in large tumors (8). Therefore, developing novel methods, specifically those based on artificial intelligence-based computational image analysis, could yield improved specificity and sensitivity of BC diagnosis and screening and earlier diagnosis of BC, finally improving the outcome and quality of life.

It has been postulated that medical images include much more information than what can be seen by the naked human eye (9). Recently, a new field of computer-aided detection (CAD), named radiomics, has been developed, which works by extracting quantitative features from medical images using a high-throughput approach (10-12). The extracted shape and textural features of radiomics are imaging biomarkers that reflect the mechanisms of cancer cells at the cellular and molecular level, which cannot be qualitatively evaluated by the human eye (13). Therefore, these data can be used as clues for the accurate diagnosis of malignancies, reduce the application of invasive diagnostic methods, improve the diagnostic accuracy and classification of the tumor, predict treatment response, and predict the survival and prognosis of the patients (14, 15). Using radiomics methods could improve the early diagnosis of BC with higher specificity and sensitivity, discrimination of benign BC masses from malignant masses, tumor molecular subtype classification, predicting of the prognosis, and optimizing the treatment method.

In this narrative review article, we aim to do a comprehensive review of the current screening and diagnostic protocols of BC, the standard radiomics workflow, and the recent progress of radiomics in the screening and diagnosis of BC based on different radiologic methods with comparison to radiologist-based imaging, the limitations and challenges confronted, and potential future directions to overcome these challenges.

## 1. Radiomics workflow *(Figure 1)*

The application of radiomics in oncology imaging requires five main steps. The first step is the acquisition of high-quality quantitative images. Standardized protocols, such as quantitative imaging biomarker alliance and quantitative imaging network (16), have been developed to improve the reproducibility and quality of the images. The second step is segmenting the region of interest (ROI), the area from which the radiomic features will be extracted. To obtain reliable data, segmentation should be reproducible and accurate. ROI segmentation can be conducted automatically, semi-automatically, or manually. Expert radiologists' manual segmentation of the ROI is the gold standard method (17, 18); however, some studies have used semi-automatic or automatic processes, which are superior if extensive data is being evaluated (19-21). Considering semi-automatic methods, two main strategies can be used for ROI segmentation: region-based and threshold-based. Threshold-based methods segment the image based on the contrasts and intensities of the images, while region-based methods work based on the voxel similarity and connection (22).

The third step is the computational extraction of features from the ROI, conducted through standard mathematical processes. These features can be divided into four different subgroups. The first group of features is shaped (morphological) features, which are 3D reconstructed images and include volume, shape, border heterogeneity, surface-to-volume ratio, sphericity, compactness, etc. An example of shape features is comparing the surface-to-volume ratio in two tumors with the same volume, of which one is speculated, and the other is round. The surface-to-volume percentage is higher in speculated tumors and thus has a higher probability of malignancy, while it is lower in round tumors and thus has a higher likelihood of being benign. The second group of features is the histogram-based features, also named first-order statistics features, commonly used for predicting patient prognosis. These features include mean, median, kurtosis, skewness, standard deviation, minimum, maximum, average volume, entropy, and uniformity (23). The third group of features includes the textural features, also named second-order statistics, which evaluate the voxels and the neighbors and commonly demonstrate intertumoral heterogeneity. Gray-level size zone matrix (GLSZM), fractal self-similarity, gray-level run-length matrix, and gray-level co-occurrence matrix (GLCM) are the standard methods that can be used to extract textural features from images. The fourth feature group is the user-selected transforms, also named higher-order statistics, including wavelet or Laplacian of Gaussian, and reflect the spatial relationships of the adjacent pixels (24). Several tools have been developed to extract the features from images, including the CGITA, 3DSlicer, Mazda, etc., which are open-access and available online (22).

The fourth step of radiomics is feature analysis. In this context, machine learning (ML) and deep learning (DL) strategies, such as random forest, principal component analysis (PCA), and most minor absolute shrinkage and selection operator (LASSO), have been extensively applied in different radiomics studies (25). The fifth and last step of the radiomics process is model building. Artificial intelligence has been widely used to obtain predictive models for the extracted features. Several methods have been used in this context, such as random forest, XG booster classifier, regularized linear regression, linear regression, and support vector machine (26, 27).

## 2. Current application of radiomics for breast cancer screening, diagnosis, and classification

Radiomics, first used for head and neck cancer imaging (28), has been recently studied for screening, diagnosis, predicting treatment response, predicting lymph node metastasis, predicting recurrence chance, and prediction of prognosis in breast cancer (29). The most studied modality for applying radiomics approaches in BC is MRI. However, other modalities have also been studied, including contrast-enhanced spectral mammography (CEST), standard mammography, US, and DBT. Studies have shown that adding radiomics to the standard radiological processing of BC can increase its diagnostic accuracy (30). Here, we aim to do a modality-based review of the current progress of radiomics for breast cancer, focusing on BC screening and diagnosis based on recently published studies, and compare AI-assisted BC screening/diagnosis with radiologist-based methods *(Table 1)*. We also discuss the potential of radiomics as the developing modality for improving the accuracy of breast cancer screening and diagnosis to decrease the bulk of invasive breast biopsies.

## 2.1. Magnetic resonance imaging (MRI)

MRI is traditionally used to screen high-risk patients with positive breast cancer gene-1, 2 (BRCA1, 2), specifically when US or mammography shows suspicious findings (31, 32). However, its role in screening intermediate and average-risk women is also being considered (33). MRI has a high sensitivity for detecting malignant breast lesions; however, since malignant and benign breast lesions share similar characteristics, it has a low specificity for BC detection (*figure 2b*) (34). Based on the evidence, MRI can successfully discriminate between malignant and benign breast lesions in up to 72% of the cases (35). Diffusion-weighted imaging (DWI) and dynamic contrast-enhancing (DCE)-MRI is the most studied MRI modality to discriminate benign from malignant breast lesions in radiomic studies (36). Most studies focused on first- and second-order statistics. The primary target lesions were BIRADS 4 and 5 for screening/diagnostic purposes. For example, in a study in 2018 that included 222 patients, the authors aimed to develop a radiomic MRI model to predict malignancy in suspicious BIRADS-4 and five breast lesions. The results showed that a kurtosis-diffusion weighted imaging model improved the sensitivity and specificity of MR-based discrimination of malignant breast lesions from benign breast lesions (37).

More recently, multiparametric MRI, combining DCE and DWI-MRI with improved specificity, has been extensively used to extract features to reduce unnecessary biopsies (38). In a study by Parekh et al., multiparametric MRI was used for texture analysis of benign and malignant breast lesions. The results indicated that the radiomic feature map (RFM) related to tumor vascular and cellular heterogeneity significantly differed between benign and malignant lesions regarding different apparent diffusion coefficients and perfusion parameters. Consistently, the multiview IsoSVM model demonstrated a sensitivity of 93%, a specificity of 85%, and an AUC of 0.91 in discriminating between benign and malignant breast lesions (39). In another study by Naranjo and colleagues, multiparametric MRI-based machine learning outperformed DWI and DCE-MRI in distinguishing benign and malignant breast lesions (40).

Although MRI is not routinely used for BC screening and diagnosis, its application could be more bright in discriminating suspicious lesions to avoid unnecessary biopsies and their cosmetic/financial burden to the patients. In a study, Pötsch et al. developed an AI classifier trained on 4D radiomic features of DCE-MRI of BIRADS-4 and five images. The AI classifier showed an AUC=0.80 for predicting malignancy in testing and 0.85 in the validation set, respectively. More importantly, further analysis demonstrated that the classifier could have reduced the rate of unnecessary biopsies up to 36.2% with a false negative score of only 4.5%. This study showed the great potential of a radiomic-based AI classifier in reducing unnecessary biopsies of suspicious BIRADS-4 and five lesions (41).

Peritumoral tissue contains essential data about the invasiveness of the tumor, and its quantitative features can be used for distinguishing benign and malignant lesions. The automatic deep learning (ResNet 50) method was trained based on 15 peritumoral tissue features in a study. The trained algorithm showed high diagnostic accuracy when the proximal peritumoral tissue's smallest bounding box was considered the input (42). Since different malignant breast lesions show other invasive behaviors, MRI-based radiomics models could also assist in identifying the phenotype/genotype of the BC, such as discrimination between TNBC from non-TNBC breast lesions (43). In a study by Ma et al., the radiomic feature-based model showed an AUC=0.741 for cross-validation and 0.867 for the testing dataset in differentiating TNBC from non-TNBC breast lesions (44).

Furthermore, the background parenchymal heterogeneity of the tumor can be identified by extracting its radiomic features and can improve the potential to discriminate breast lesions and specifically define TNBC (45). In a study, quantitative texture features of background parenchyma were extracted to identify the TNBC from other types of BC. The results indicated that radiomic features of the lesion had an AUC=0.782 in discriminating TNBC from non-TNBC tumors. When the features of the background parenchymal features were added, the AUC was significantly improved to 0.878 (46). Similar results were obtained in another study that combined breast-parenchymal enhancement (BPE) features with tumor-extracted features (45).

Considering BC screening and diagnosis, comparing the AI-based model's diagnostic performance with radiologists' performance could be attractive. In a study in 2018 by Truhn et al., a convolutional neural network (CNN)-based model outperformed radiomics in discriminating benign and malignant breast lesions. However, radiologists' performance was superior compared to both of them (47). In a multicenter study, Naranjo et al. compared the performance of a radiomic-based machine learning decision-making model with the performance of two radiologists in discriminating between benign and malignant lesions. The performance of the multiparametric MRI radiomics with apparent diffusion coefficient (ADC) and BIRADS was similar to the implementation of radiologists in discriminating benign from malignant lesions (p=0.39). Thus, the authors concluded that this method could assist the decision-making of less experienced radiologists in differentiating between benign and malignant lesions (34). In another study, the authors developed a machine learning model to evaluate the assistance of radiomic-based machine learning models to improve the diagnostic performance of radiologists. The study's results demonstrated that combining the multiparametric MRI-based DL model improved the diagnostic performance of two junior radiologists, from an AUC=0.823 to 0.876 and 0.833 to 0.885, respectively. However, the application of this model achieved comparable results to that of senior radiologists' performance with an AUC=0.944 in discriminating TNBC from fibroadenoma lesions (48).

## 2.2. Mammography *(Figure 3, Figure 4a, b)*

Mammography is the most used diagnostic modality applied for BC screening. Architectural distortions, microcalcifications, and tissue asymmetry are the most important biomarkers for BC diagnosis in mammography (49, 50). Based on different studies, the sensitivity of mammography for BC screening is 75-90% (51). Its specificity for BC screening is consistently 80-90% (52, 53). Although adjunctive ultrasonography and mammography can improve the accuracy of BC screening, specifically in small breast lesions under 2cm (54, 55), novel radiomics methods could improve its accuracy more efficiently.

Furthermore, mammography's overall sensitivity and specificity for BC diagnosis is around 90%; however, a positive predictive value of 15% has been reported for mammography (56). Several studies have focused on improving the accuracy of BC diagnosis using AI-based radiomics approaches (57, 58). According to many mammography images for BC screening in the clinical setting, applying AI-based algorithms to mammography can reduce the burden on radiologists and improve diagnostic accuracy. In this regard, several studies have developed AI algorithms based on deep neural networks in recent years (59-66). An example is a large-scale study conducted by Kooi et al. on 45000 mammography samples. This study evaluated two approaches: a convolutional neural network (CNN) and the other performing by manual feature setting. The results demonstrated that the fully automated CNN outperformed conventional CAD methods in diagnosing BC at high and low sensitivity modes (67).

Combining trained AI-based deep learning tools with radiologist-based assessments could provide higher accuracy in identifying BC. In a study in 2020, Schaffter et al. aimed to evaluate whether combining radiomics with radiologist assessment can improve the accuracy of mammography-based BC screening. To do this, they collected data from 144231 mammograms to train and validate a deep learning algorithm. The analysis revealed that the combination of this AI-based diagnosing modality with the radiologist assessments had an AUC= 0.942, higher than the AI-based and radiologist-based assessments. Moreover, the combination method also showed a higher specificity of 92%. The authors concluded that although the machine learning algorithms were inferior to radiologist assessments, combining these methods with a single-reader radiologist assessment improved the accuracy of mammography interpretations for BC screening (68).

In 2018, a retrospective study by Mao et al. aimed to develop radiomic features for diagnosing BC using mammography images. Among different regression models, the logistic regression model showed better results, having sensitivity, specificity, and diagnostic accuracy of 0.983, 0.975, and 0.978 for training data and 0.867, 0.900, and 0.886 for testing data, respectively. The authors concluded that quantitative mammography imaging could be an excellent diagnostic approach for BC diagnosis (69). In a retrospective study by Rodriguez et al., the authors aimed to evaluate the diagnostic efficacy of an AI-based system for detecting malignancy in 2652 separate exams from seven different countries interpreted by 101 radiologists. The results of this study indicated that compared to radiologists (AUC=0.914), the function of the AI-based system (AUC=0.814) was not inferior.

Moreover, the AI-based system did better than 61.4% of the radiologists in diagnosing the malignancy, meaning that a high-performance AI system can do better than an average radiologist in diagnosing BC in the clinical setting (70). Another paper by the same authors evaluated the feasibility of reducing the workload of mammographic studies using AI-based systems to exclude the lower malignancy likelihood results. The results of this study indicated that setting the AI threshold of malignancy probability at 2 and 5 reduces the mammography-reading workload up to 17% and 47%, respectively. However, choosing 2 and 5 points was also associated with excluding 1% and 7% of actual positive exams, respectively. This study showed that pre-selection of mammography exams using an AI-based system could reduce radiologists' screening mammography reading tasks by excluding the low-likelihood exams. Nevertheless, further studies are still indicated to define the best likelihood threshold at which the lowest likelihood exams are excluded, but the true-positive studies are not excluded (71).

A recently published meta-analysis study aimed to evaluate the diagnostic accuracy of current machine learning methods on mammography. Thirty-six studies were included, of which the overall specificity, sensitivity, and AUC were 0.84, 0.83, and 0.90 (72). Nevertheless, since the authors indicated that Deek's linear regression model showed a bias, the data of this meta-analysis should be cautiously interpreted. Another systematic review and meta-analysis study was conducted, including 19 studies that had used radiomics as the preoperative diagnostic method for BC. They found that radiomic models had an AUC of 0.91, specificity of 0.83, and sensitivity of 0.84 in detecting BC in the preoperative setting (73). However, since the number of primary studies included in the meta-analysis was relatively small, this study's results must be investigated by large-scale trials.

One of the essential goals of BC screening by mammography is to reduce the number of unnecessary biopsies in high-risk patients. Mammography-based radiomics have shown promising results in this area. An example is a study conducted by Drukker et al. that combined the quantitative three-compartment breast (3CB) image analysis with mammography radiomics to assess its potential to reduce unnecessary biopsy sampling. The results indicated that compared to conventional mammography, the combined method could reduce 35.3% of total biopsies and had a sensitivity of 97% (74). Microcalcifications, especially when they are present in dense breast parenchyma, could be challenging to distinguish and, therefore, may lead to unnecessary biopsies. Applying radiomics to mammography images could improve detection accuracy and reduce unnecessary biopsies. In a study, texture analysis of the surrounding tissue of the microcalcifications (excluding microcalcifications) had an AUC of 0.96 and reduced unnecessary biopsies (75). Another study by Wang et al. have shown similar results (57).

A significant barrier to mammography-based BC screening is high densities, which can mask the tumor and delay diagnosis. Density is commonly evaluated based on the BI-RADS density score. A study aimed to develop a texture-based model for predicting the risk of tumor masking in high-density mammograms. The extracted model had an AUC of 0.75, while volumetric breast density, age-adjusted BI-RADS density, and BI-RADS density had an AUC of 0.72, 0.71, and 0.64, respectively. The authors concluded that the radiomic texture metrics could predict the masking of the breast lesion more accurately than other prediction methods (76). The features of the contralateral breast parenchyma hold important histopathological data which could improve the radiomic-based diagnosis of BC. In a retrospective study, Li et al. included mammography data from 182 patients, comparing the radiomics features of the lesions with their contralateral ROI breast parenchyma. Six radiomic features were selected, including skewness, speculation, circularity, margin sharpness, size, and power-law beta. The results of this study indicated that the combination of parenchyma and lesion classifier (AUC=0.84 ± 0.03) did better than the lesion classifier alone (AUC=0.79 ± 0.03) (77).

## 2.3. Contrast-enhanced spectral mammography (CESM)

Contrast-enhanced spectral mammography (CESM) is a developing method for BC diagnosis with high sensitivity after intravenous (IV) injection of contrast material (*Figure 2a)* (78, 79). Recently, CESM has shown similar accuracy to MRI in discriminating breast lesions (80, 81). Previous studies have focused on utilizing radiomics-based methods for analyzing CESM in differentiating malignant from benign breast lesions (82). In a study, Massafra et al. studied 464 features from CESM exams of 53 patients. They trained three different classifiers, including logistic regression, naïve Bayes, and random forest, on each subset of principal components, of which the random forest-based classifier showed the best potential to predict the malignancy of the ROIs with a sensitivity of 88.37% and a specificity of 100% (83). Radiomics feature selection using CESM can also differentiate invasive forms from non-invasive forms of breast cancer. In a study, six different radiomic features were analyzed in terms of their power to discriminate the invasiveness of the BC. It was shown that the co-occurrence matrix (COM), in combination with the first-order histogram (HIS) or mutual information (MI) coefficient had an accuracy of 87.4% for differentiation of invasive from non-invasive BC (84). In another study, the authors compared the radiomics analysis of DCE-MRI and CEM in evaluating BC invasiveness. This retrospective analysis included data from 48 women with 49 biopsy-confirmed BC. Analysis of DCE-MRI radiomics showed an accuracy of 90% in discriminating invasive from non-invasive BCs, while it was 92% for CEM (85). One of the challenges of radiographic diagnosis of BC is the small size of lesions, especially <1cm. Radiomics could also help detect small malignant lesions. In 2020, Lin et al. conducted a retrospective study on 139 patients with lesions of sub-1cm diameter. Their study showed that a radiologic nomogram combined with CES-based radiomic features and predictive factors of age and BI-RADS score had an AUC=0.94 in identifying benign from malignant breast lesions of <1cm (86).

The perilesional area of the tumor contains essential information that could help diagnose malignant lesions. A retrospective study by Wang et al. aimed to analyze the radiomic features of the perilesional regions of 190 women with 223 breast lesions. They surveyed 4098 radiomic characteristics from 7 ROIs. Among the methods used, radiomic features of the annular perilesional region of 3mm had the highest AUC=0.93 for distinguishing benign from malignant lesions. Moreover, the authors reported that combining the radiomic analysis of the ROI with the annular perilesional region of 3mm had a higher AUC=0.94; thus, the combinatory method could do better in diagnosing BC (87).

It is already distinguished that some factors, such as lesion size, can influence the interpretation of data derived from CESM (88). To determine the factors influencing the understanding of radiomic features extracted from CESM, Sun et al. aimed to identify the factors that influence the classification of CESM-based radiomic models in benign and malignant breast lesions. To do this, they collected the misclassified data by the random forest (RF) algorithm and Least absolute shrinkage and selection operator (LASSO) regression models. Multivariate analysis demonstrated that the presence of air-trapping infarcts and the small size of the lesion were the two factors that led to misinterpretation of the malignant breast lesions. However, considering misclassifications of benign breast lesions, the presence of ripple and/or rim artifacts and larger lesion size were the two crucial influencing factors (89).

Combining CESM radiomic feature analysis with another diagnostic method could improve the accuracy of benign/malignant lesion discrimination. In a study by Fusco et al., 54 patients with 79 histopathologically proven breast cancers underwent CEM and DCE-MRI and were analyzed for 48 textural radiomic features. Considering CEM, skewness (AUC=0.71) and kurtosis (AUC=0.71) were the best predictors of malignancy. Consistently, considering components extracted from DCE-MRI, gray-level run-length matrix (AUC = 0.72), GLN (gray-level non-uniformity) (AUC = 0.72), entropy (AUC = 0.70), energy (AUC = 0.72), and RANGE (AUC = 0.72) were the best predictors of malignancy (90).

## 2.4. Digital breast tomosynthesis (DBT) *(Figure 4 c, d)*

Mammography has low sensitivity to detect breast lesions in dense breasts due to the overlapping of the tissues. DBT is the pseudo-3D reconstruction of mammographic images that enhances the discrimination of breast tissues (91-94). Therefore, one of the applications of DBT is to diagnose malignancy in breasts with dense tissue (95). Recently, AI-based radiomics approaches have been used in DBT to improve diagnostic accuracy (96, 97). In 2018, Tgliafico et al. included two groups of patients, 20 cancer-detected exams with dense breasts and negative mammography, with exams from 20 healthy-matched individuals. Three radiomic features correlated with tumor size, including 90 percentile, skewness, and entropy.

Moreover, entropy was related to the estrogen receptor status of the BC (98). In another study by Sakai et al., using machine learning and radiomic features, an automated classifier was trained to distinguish benign from malignant breast lesions on DBT. Among different classifiers, the accuracy of the support vector machine-based classifier was 55%for benign lesions and 84% for malignant lesions (99). In a study that compared the performance of a CNN for the classification of 76 breast lesions, DBT did better in the classification of architectural distortions of malignant masses than mammography (100).

Data extracted from the peritumoral area are essential for distinguishing benign and malignant lesions in DBT. In a recently published study, Niu et al. aimed to extract the DBT-based peritumoral radiomic features for differentiating benign lesions from malignant ones. They extracted radiomic features from the lesion and the 2mm diameter of the peritumoral area. The radiographic nomogram incorporated menstruation status, age, and radimoic features, which showed a specificity of 0.946, sensitivity of 0.970, and AUC of 0.980 in the training set. However, it had a specificity of 0.966, a sensitivity of 0.909, and an AUC of 0.985 in the validation cohort. The authors concluded that the radiomic nomogram that integrates the clinical data (menstruation status and age) with peritumoral radiomic features could help increase the accuracy of BC diagnosis (101).

Microcalcification clusters can disturb the diagnosis of BC in DBT images because of their spanning across slices. Therefore, differentiating malignant and benign microcalcification clusters in DBT is a critical issue radiomics studies address (102). In a study, the authors developed a semi-automatic segmentation radiomic-based approach for differentiating benign microcalcifications clusters from malignant ones. Among different features, the view-based mode and case-based mode of both 2D and 3D-domain radiomis had the highest AUC of 0.834 and 0.868, respectively. Therefore, the authors concluded that radiomic-based models could perform well in differentiating microcalcification clusters that are challenging to diagnose on DBT (103). In another study, to reduce the size of false positive calcifications in DBT, the authors developed a decision support system based on radiomic classifiers combined with the BI-RADS scoring system. The trained machine learning classifier reduced the false positive to half and improved the positive predictive value up to 50% (104). Consistently, Xiao et al. proposed an ensemble convolutional neural network (CNN) consisting of a 2D ResNet34g for extracting features from 2D slices and a 3D ResNet to extract 3D contextual features. To avoid the noises of the anisotropy, 3D ResNet was built using anisotropic 3D convolution. The results of this study indicated that 2D ResNet34 had an AUC=0.8264, anisotropic 3D ResNet had an AUC=0.8455, and deep learning ensemble strategy had an AUC=0.8837. The authors concluded that by mixing the 2D and 3D features, the ensemble CNN had improved performance and reduced false positive results according to 0.0435 increase in AUC (105).

## 2.5. Ultrasound imaging (US) *(Figure 5)*

Ultrasound imaging (US) is integral to the BC screening system, specifically in dense breasts. However, it has low specificity, increasing the rate of false-positive results and burdens unnecessary biopsies. Therefore, quantitative radiomics-based approaches have been widely studied to improve BC's early diagnosis and classification and reduce unnecessary biopsies (106-112). It has been illustrated that histogram, texture, and shape-oriented features are the essential gradients of US images to distinguish between benign and malignant lesions (113). In a recent study by Hassanien and colleagues, the authors proposed a deep learning-based radiomics method on ultrasound sequences of the breast. They used a deep learning network, ConvNeXt, for radiomics feature extraction and a pooling mechanism for generating malignancy scores for every sequence. The authors showed that the proposed ConvNeXt method outperformed all previously developed methods, having an accuracy of 91%. Since the quality of the US images affects the accuracy of the malignancy-predicting models, ignoring the low-quality US images improved the score pooling model's accuracy (114). In another recent study, Jabeen and colleagues proposed a new CNN model for predicting malignancy in US images. When validated on the Breast Ultrasound Images (BUSI) dataset, this model achieved the highest accuracy of 99.1% (115). Appropriate support medical decision-making models based in the US could improve the classification of breast lesions and reduce the rate of benign lesion biopsy. Interlenghi et al. proposed a machine learning model for predicting BI-RADS scores based on the US. The proposed model reduced the biopsy rate from 18% to 15% in benign lesions. The external validation of the model on two different datasets had a positive predictive value (PPV) of 45.9% and 98% sensitivity, compared to the 41.5% PPV of a radiologist ($p<0.005$) in the first dataset. Validation testing on the second dataset showed a PPV of 50.5% and sensitivity of 94%, compared to the PPV of 47.8% of a radiologist ($p<0.005$). Moreover, compared to a board-certified breast radiologist, in six of nine images, the model outperformed the radiologist in giving a lower BIRADS score to benign masses (116). In another study, the proposed model reduced 67.86% of the unnecessary biopsies; however, a false-negative rate of 10.4% was observed (108).

Deep learning models based on features extracted from morphological characteristics on US imaging can also distinguish the type of breast cancer subtype. In a study, morphological characteristics of US images of 282 women with breast cancer were included. While the TNBC subtype tended to have a parallel orientation and more round/oval morphology in the US, the luminal A subtype had a more distinctive feature but did not show a similar direction. Moreover, tumor size was associated with hypoechogenicity, a non-circumscribed margin, and irregular shape (117).

Ten radiomic features were extracted from 206 percutaneously biopsied lesions (62 malignant and 144 benign) in another retrospective study. Different methods were used, of which the support vector machine (SVM) had 76.9% specificity, 71.4% sensitivity, and the highest AUC of 0.840 (118). Data from US imaging can also improve the differentiation of benign fibroadenoma from triple-negative breast cancer (TNBC) (97, 119-122). A study by Du et al. developed a nomogram to differentiate TNBC from fibroadenoma based on radiomics features extracted from the US and patient clinical data. In the training cohort and validation, the nomogram had an AUC of 0.986 and 0.977, respectively, outperforming clinical models and radiomics signatures (123). In a study to differentiate between phyllodes tumor among fibroepithelial lesions in the US, the authors extracted 93 radiomic features from 182 fibroepithelial lesions biopsied and diagnosed via core needle biopsy. In the validation set, the radiomics-based classifier had an AUC of 0.765, an accuracy of 0.703, a specificity of 0.5, and a sensitivity of 0.857 (124).

Nomograms combining radiomic results with data from the BI-RADS scoring system could provide additional important information on differentiating malignant lesions from benign lesions. In a study in 2019, Luo et al. developed a nomogram based on US BIRADS 4 or 5 lesions and nine radiomic features of the US images. Of the 315 pathologically-proven samples, 211 were included in the training and 104 in the validation groups. The nomogram combining radiomic features and BIRADS scoring showed a better discrimination power with an AUC=0.928 than radiomics and BIRADS scoring alone (p value=0.029 and 0.011, respectively). Therefore, the authors concluded that the nomogram incorporating radiomics feature with the BIRADS category is a potentially useful method for predicting malignancy in BIRADS 4 and 5 US lesions (125).

Tumor vasculature consists of essential data for differentiating malignant and benign lesions. Color Doppler sonography is used for assessing the vascularity of breast lesions. Higher vascularity lesions are more suspicious of being malignant. In 2020, Moustafa et al. extracted quantitative radiomic features from the color Doppler ultrasound images to develop a diagnostic model for BC diagnosis. To establish an AdaBoost ensemble classifier, two Doppler and seven grayscale features were extracted from 159 Doppler exams combined with a BIRADSUS category and age information. Training the model based on color Doppler and grayscale features improved the AUC from 0.925 to 0.958, reduced borderline diagnosis, and enhanced the diagnostic performance (126). In another study, Ternifi et al. developed a quantitative non-contrast high-definition microvasculature imaging (HDMI) for breast cancer detection in highly-suspicious breast masses. Four tumor microvessel features, including spatial vascularity pattern (SVP), bifurcation angle (BA), Murray's deviation (MD), and microvessel fractal dimension (mvFD), were considered. For lesions greater than 20mm, the HDMI model had a sensitivity of 95.6 and a specificity of 100%. Adding the BIRADS score to the HDMI biomarkers produced a prediction model with 89.2% specificity, 93.8% sensitivity, and an AUC of 0.97 (127).

Similar to other modalities, peritumoral data from the US could also provide additional information that could improve the classification of breast lesions. Fuse the info of multiple tumoral-regions (FMRNet) is an example of a convolutional neural network developed for breast tumor classification by combining the radiomic signature feature from multiple tumoral areas. When tested on the UDIAT dataset, FMRNet had an accuracy and specificity of 0.945 (128).

# 3. Limitations, challenges, and future directions

Radiomics has been extensively applied in BC screening and diagnosis. Despite progress in this area, no clinical application is available for radiomics, and we still have a long way to bring radiomics to the bedside (129). Based on a study, it has been shown that radiomic studies have an overall insufficient scientific quality and radiomics reporting, specifically in the fields of open science category, clinical utility, and reproducibility. Moreover, this retrospective meta-analysis study evaluated the quality of radiomic studies via radiomics quality scoring (RQS). The results indicated a low rate of the 17 included studies (130). This section discusses the limitations and challenges of radiomics application in BC screening and diagnosis and possible solutions.

The first and significant challenge is the reproducibility of the features (131). The variability of radiomic features, the heterogeneity of imaging protocols (especially in MRI), the heterogeneity of processing methods, and the retrospective nature of the studies are the significant factors that dampen the reproducibility of radiomics studies (132). It has been shown that voxel shape and slice thickness are the two features that have the most effect on the reproducibility of elements (133). Another major factor that affects the reproducibility of radiomics is the method used for ROI segmentation. Although automated methods are suitable for extensive data, they can only explore their best function if the lesions have a well-defined margin (134).

Moreover, although the manual method is more accurate, it is time-consuming, and the intra and inter-interpreter variability reduces the reproducibility of the radiomics method (135). The application of novel CNN-based methods, such as the Ensemble Deep Neural Network model, has contributed significantly to automating the segmentation task, as shown in a recently published study (136). Another critical factor affecting the reproducibility of high-level features is the image quality, which has affected US-based radiomics studies more than other modalities (137). Augmentation of the images using deep learning models can help gain better results. One of the methods used for image augmentation is the generative adversarial network (GAN) (138).

A study used GAN to generate high-quality US images of breast lesions. Using a CNN model for interpretation, synthetic GAN-augmented images resulted in better classification of malignant from benign breast lesions compared to the baseline method, with an accuracy of 90.41% (139). Developing novel technologies and algorithms for high-quality image acquisition could yield better results and higher accuracy in radiomic studies shortly. Examples are novel MRI-based developing technologies, such as blood oxygen level-dependent MRI, ultrafast MRI, abbreviated MRI, and arterial-spin labeling (ASL)-MRI, which produce high-quality images for feature extraction (33, 140, 141). The scanner model, imaging parameters, and imaging protocol heterogeneity are among other factors that may influence the result and reproducibility of the radiomics methods, specifically regarding MRI, which includes many more parameters than US or mammography (129). An example of the scanner effect is a study that showed that compared to breast tumors, the radiomic features of the fibroglandular tissue are more dependent on the MRI scanner model (142-144).

Moreover, the same algorithm may show different results in two sample sizes. Therefore, the sample size is another issue that can influence the results of a study. From a statistics view, predictive models built on a more extensive study size have more accuracy ref. Thus, since most radiomic studies use a small size, their results must be externally validated on larger datasets.

In addition, variations in the imaging protocols and image post-processing for image normalization are two critical factors that influence the quality of the images and can thus reduce the reproducibility of BC radiomic studies. Proper normalization protocols are required to minimize the effects of these factors on database quality. Nevertheless, the post-processing algorithms (such as the black box) are not publicly accessible (145). Image acquisition via standardized methods can reduce the heterogeneity of images. In training the algorithms, the hardware used, dataset size and type, dataset quality, and data architecture are essential factors affecting the results. To reduce the time required for training the algorithms, parallel computation strategies on graphics processing units (GPUs) can be used (146).

Additionally, overfitting and validation are essential barriers in radiomics studies due to the extensive application of machine learning algorithms. To solve this problem, using multicenter data could be a possible solution. Since multicentric studies might not be available for all, the best solution might be the application of open databases. Dataset accessibility is another solution with an essential effect on the external validation and reproducibility of radiomics studies. Limited dataset sharing and availability of the algorithms used in different studies is one of the critical challenges. The cancer imaging archive (TCIA) is the most used open database for radiomics studies; however, regarding breast cancer, it only includes 1000 mammography and 500 breast MRI images, which is not sufficient (147). Regarding mammography, several other public datasets have been recently available. These datasets include the BGH, DDSM, MIAS, INBreast, OPTIMAM, VICTRE, BCDR, and LAPIMO (148).

Several factors can affect the results of discriminating benign and malignant lesions by radiomics studies. The first issue is the modality used for imaging. The same algorithm could demonstrate different performances when applied to distinct modalities. An example is a study conducted by Antropova et al. in 2017 which applied a CAD method that exploited handcrafted features with a pre-trained CNN model for BC diagnosis. The model was examined on mammography, US, and DCE-MRI. This method demonstrated an AUC of 0.89 for DCE-MRI, 0.86 for mammography, and 0.90 for the US (149). Intra and inter-observer segmentation variability is another issue that can affect feature extraction results. Automatic segmentation of images could reduce the effects of this variability. However, a study showed that 32.8% of Pyradiomics and 41.6% of radiomiX features were robust, meaning that inter-observer variability did not affect them (135).

The progressing role of deep learning in radiomics has resulted in the development of fully automated models that can conduct all phases of radiomics workflow, from image segmentation to classification (150). An example is a recent work performed by Vigil et al., which developed a dual-intended deep-learning model that showed an accuracy of 78.5% in the classification of sonographic breast lesions (151). Several other models have been developed that focus on classifying breast cancer, and most of them are surveyed on currently available online databases.

To improve the accuracy of screening BC, pre-screening risk assessment with thermal radiomics and AI can be used to generate a personalized risk score for BC. Thermalytix Risk Score (TRS) is an AI-based, radiation-free, non-invasive method for identifying high-risk populations based on the metabolic activity of breast tissue. Compared to age-based risk assessment, this method has shown a 21% improvement in AUC (3). With the development of AI and deep learning strategies, novel methods incorporating the potential of AI may be used for radiation-free pre-screening personalized risk assessment of patients. This method could finally lead to higher accuracy in breast cancer screening (152, 153). Another topic of significant interest is developing automatic predicting models based on transfer learning from neural network (NN) systems, which has gained much attention recently (154). It has been shown that NN-based models are more accurate than conventional machine/deep learning tools such as SVR/KNN; however, NNs are more expensive and have more complexities (5, 155-157). Currently, several CNN-based models have been developed and tested on different modalities, mostly mammogram databases which have shown the high accuracy of these models in discriminating benign from malignant breast lesions (158, 159).

Another developing approach is developing nomograms for obtaining a more accurate diagnosis in BC studies. The concept is that different factors, such as age and menstrual cycle, can affect the density of the breast. Therefore, to obtain more reliable and accurate results, recent studies have incorporated radiomic studies with patient clinical information, such as age, menstrual cycle, and BI-RADS score, to develop more accurate nomograms. Nomograms have shown better accuracy and higher AUCs in predicting breast cancer in different studies (160-164) and should be considered in further studies.

The accuracy of imaging studies for small lesions, specifically those less than 1cm, is low. Therefore, one of the main challenges would be the identification of small-size lesions by radiomics features. Despite the progress in improving the diagnostic accuracy of radiomics for subcentimeter lesions, future studies are still indicated to achieve more accurate models and methods (165). Another significant issue is the undetermined zone, where benign and malignant tissues share the same radiomic features, accounting for 15% of lesions. The most common lesions of the undetermined zone are Fibroadenoma, phyllodes tumor, etc. An undetermined radiomics zone is a dilemma in the interpretation of radiomic studies. Despite studies showing the potential of computer-aided feature extraction to replicate the human BIRADS score definition (166), currently, the radiomic features can not discriminate the undetermined radiomic lesion. Therefore, expert radiologists' interpretation of the data is still required to discriminate these lesions by defining the BI-RADS score (167).

False-positive results, specifically in MRI studies, are a significant issue that requires specific attention since they may result in unnecessary biopsy sampling in benign lesions. To reduce false-positive results of radiomic features extracted from DCE-MRI, Zhao et al. combined radiomic features extracted from mammography with features extracted from DCE-MRI to develop a combined machine learning-based model for BC diagnosis. Compared to the accuracy of radiomic features of DCE-MRI alone, which had an accuracy of 78.8% and specificity of 69.6%, the combined DCE-MRI and mammography-based model had an improved accuracy (83.3%) and specificity (82.1%). Studies focusing on combining other modalities with MRI findings could improve the accuracy of radiomic-based models and reduce false-negative results in the future.

DBT acquisition angle is another factor that can impact the quality of breast parenchymal feature findings in DBT. The effects of acquisition angles of 15$^{o}$ and 40$^{o}$ on radiomic-based BC diagnosis were analyzed in a study. GLCM features significantly differed between 15$^{o}$ and 40$^{o}$ images, while there was no difference in histogram-based features. The texture analysis revealed that the DBT acquisition angle affects the results of radiomics studies, especially GLCM (168).

Besides screening and diagnosis, recent studies have also focused on extracting radiomic features for predicting other aspects of BC, including classification of its molecular subtypes (147, 169-173), axillary lymph node metastasis (174), treatment response (175, 176), prediction of prognosis (143, 177), prediction of survival (178), and prediction of its risk of recurrence (179). The growing role of radiomics and AI in this area may assist patient management and contribute to the personalized treatment of patients shortly (180). The promising results of current studies indicate the potential of radiomics to become part of the routine decision-making for clinical practice regarding BC (181).

# Conclusion and discussion

In recent years, radiomics, combined with the expanding role of machine learning and AI, has shown great potential in breast cancer research, specifically in improving breast cancer screening and diagnostic accuracy. This review article presents the current progress of radiomic-based studies for BC screening and diagnosis using MRI, US, DBT, mammography, and CESM modalities. MRI is getting more attention in BC screening, especially after recent studies showing its potential to reduce unnecessary biopsies (41). However, MRI-based radiomics can not replace a biopsy's role in diagnosing BC, at least in the near future (182). Moreover, based on evidence, combining radiomic features of different modalities can improve the accuracy of diagnosis and reduce false-positive results. The big question is whether radiomic-based AI models can replicate the role of radiologists in diagnosing BC. Despite promising results of radiomic-based models' performance, in most cases, these models did not perform better than expert radiologists (183). However, radiomic-based machine learning models did improve the performance of less-experienced radiologists. Therefore, the development of radiomic-based models could improve the decision-making capability of radiologists in BC diagnosis. Moreover, the growing role of AI, deep learning, machine learning, and convolutional neural network (CNN) models in radiomics (184, 185) could improve the accuracy of radiomics and enhance its application in the screening and diagnosis of BC in the clinical setting.

# Declarations

## Conflicts of interest

The authors declare that the research was conducted without commercial or financial relationships, which could be considered a potential conflict of interest.

## Ethics approval

Since this article is a review manuscript and no specific patient or animal material has been used, the ethics approval is not applicable for this study.

## Informed consent

Not applicable

## Data availability Statement

Data will be available upon request by reviewers.

| Author | Radiomic features | Specificity and Sensitivity | AUC | Findings |
|---|---|---|---|---|
| Bickelhaupt et al. (37) | Morphological features | 74.2% and 91.8% | 0.91 | A kurtosis-diffusion weighted imaging model improved the sensitivity and specificity of MR-based discrimination of malignant breast lesions from benign breast lesions. |
| Parekh et al. (38) | First and second-order Radiomics features | 80.5% and 82.5% | 0.87 | Extracting Radiomics features using the Multiparametric imaging Radiomics (mpRad) framework demonstrated higher AUC (9-28) than single Radiomics parameters. |
| Zhou et al. (42) | 99 features were calculated and 15 were used for training the algorithm | N/A | N/A | ResNet 50 method was trained based on 15 peritumoral tissue features. The trained algorithm showed high diagnostic accuracy when the smallest bounding box of the proximal peritumoral tissue was considered as the input. |
| Parekh et al. (39) | First-order statistics and gray level co-occurrence matrix | 85% and 93% | 0.91 | The multiview IsoSVM model demonstrated a specificity of 85%, a sensitivity of 93%, and an AUC of 0.91 in classifying benign and malignant lesions. |
| Hoa et al. (186) | 1046 features | N/A | 0.71 for T1+C, 0.69 for T2, and 0.77 for T1+C+T2 features | The fusion model acquired an accuracy of 74% for predicting benign and malignant lesions of suspicious BIRADS4 findings of contralateral breasts. |
| Ma et al. (44) | 15 features | N/A | 0.741 for cross-validation and 0.867 for the testing dataset | The Radiomics model based on 15 features had an AUC=0.741 for cross-validation and 0.867 for the testing dataset. |
| Naranjo et al. (40) | 5 features | N/A | 0.79 for DWI-extracted features, .83 for DCE-based features, 0.85 for | Machine learning using multiparametric MRI had the highest AUC compared to DWI and DCE-extracted features (AUC=0.85) with a diagnostic accuracy of 81.7%. |
| | | | multiparametric MRI | |
| Whitney et al. (166) | 38 features | N/A | 0.728 | Radiomics features outperformed size alone in discriminating benign breast lesions from luminal type A lesions. |
| Yin et al. (48) | N/A | N/A | 0.944 | The combination of the multiparametric MRI-based DL model improved the diagnostic performance of two junior radiologists, from an AUC=0.823 to 0.876 and 0.833 to 0.885, respectively. The model was comparable to senior radiologists' performance (p<0.05). |
| Wang et al. (46) | 85 features | N/A | 0.782 without background parenchymal features, and 0.878 with background parenchymal features | Radiomics features had an AUC=0.782 in discriminating TNBC from non-TNBC lesions. When the features of the background parenchymal features were added, the AUC was significantly improved to 0.878. |
| Agner et al. (43) | N/A | N/A | 0.73 | SVM classifier had an AUC=0.73 in discriminating TNBC from no-TNBC lesions. |
| Militello et al. (187) | 107 features | 0.741 and 0.709 | 0.725 | The model combining SVM and UDFS had an NPV=0.75 and PPV=0.72 in the blind test set. |
| Pötsch et al. (41) | 86 features | N/A | 0.80 in the testing set and 0.85 in the validation set | The AI classifier showed an AUC=0.80 for predicting malignancy in testing and 0.85 in the validation set, respectively. More importantly, the classifier could reduce the rate of unnecessary biopsies up to 36.2% with a false negative score of only 4.5%. |
| Naranjo et al. (34) | N/A | N/A | multiparametric Radiomics with DWI score and BI-RADS=0.93, multiparametric | The performance of the multiparametric MRI Radiomics with apparent diffusion coefficient and BIRADS was similar to the performance of radiologists in discriminating benign from malignant lesions (p=0.39). |
| | | | Radiomics with ADC values and BI-RADS=0.96 | |
| Antropova et al. (149) | N/A | N/A | 0.89 for DCE-MRI, 0.86 for mammography, and 0.90 for the US | The proposed $CAD_x$ methodology had an 0.89 for DCE-MRI, 0.86 for mammography, and 0.90 for the US in detecting benign from malignant breast lesions. |
| Gullo et al. (188) | SZM-based gray level variance, first-order coefficient of variation, GLCM | 91.4% and 63.2% | N/A | The machine learning and Radiomics coupled methodology had a PPV of 82%. Sensitivity of 63.2%, specificity of 91.4%, and diagnostic accuracy of 81.4%.in benign/malignant classification of subcentimeter lesions. |
| Yang et al. (45) | 18 features | N/A | The model produced from Radiomics features extracted from the tumor had an AUC=0.865, which increased to 0.919 after fusing the breast parenchymal enhancement (BPE) features. | BPE features supply complementary information to tumor-extracted features and therefore their addition improved the diagnostic accuracy of DCE-MRI-based Radiomics. |
| Zhao et al. (189) | N/A | Specificity of DCE-MRI extracted features alone=69.6%, specificity of combined mammography and DCE-MRI extracted | N/A | Compared to the accuracy of radiomic features of DCE-MRI alone which had an accuracy of 78.8% and specificity of 69.6%, the combined DCE-MRI and mammography-based model had an improved accuracy (83.3%) and specificity (82.1%). Therefore, the machine-learning model built from combined mammography and MRI features can reduce false-positive results of MRI. |
| | | features=83.3% | | |
| Verburg et al. (190) | 49 features | N/A | N/A | CAD reduced false-positive results of BIRADS-3 and 4 lesions on multiparametric MRI. |
| Schaffter et al. (68) | N/A | 92% | 0.942 | AI-based modality combined with the radiologist assessments had an improved AUC= 0.942, which was higher than both the AI-based and radiologist-based assessments |
| Wang et al. (191) | 1370 features | 0.778 and 0.950 | 0.87 for the machine learning model, 0.90 for the nomogram | The nomogram incorporating clinical data to the radiomic features had a significantly improved AUC (P<0.001). |
| Wang et al. (57) | 188 features | 60% and 98.8% in the training set, 66.7% and 95.8% in the test set | 0.934 in the training set, 0.901 in the test set | The deep learning model based on radiomics features extracted from mammography images may reduce the need for unnecessary biopsies of suspicious BIRADS-4 and 5 lesions. |
| Li et al. (77) | Power law beta, margin sharpness, size, circularity from the tumor feature set, skewness, and speculation | N/A | Combination of parenchyma and lesion classifier (AUC=0.84 ± 0.03), lesion classifier alone (AUC=0.79 ± 0.03) | The combination of parenchyma and lesion classifier (AUC=0.84 ± 0.03) did better than the lesion classifier alone (AUC=0.79 ± 0.03) for radiomic-based discrimination of benign from malignant breast lesions. |
| Rodriguez et al. (70) | N/A | N/A | 0.84 | The performance of the AI-based system was non-inferior to the performance of 101 radiologists. The AI-based system did better than 61.4% of the radiologists. |
| Lei et al. (192) | 8286 features | N/A | 0.80 | A nomogram incorporating six radiomic features with the menstrual status reached an AUC=0.80 in discriminating malignant and benign BIRADS 4 calcification. |
| Mao et al. | 51 features | 0.975 and 0.983 in | N/A | The logistic regression model can provide |
| (69) | | the trading set, 0.900, and 0.867 in the test set | | complementary findings to radiologists. |
| Huynh et al. (59) | N/A | N/A | N/A | CNN-based extracted feature classifier's performance was comparable to analytically extracted feature classifier. Moreover, the transfer learning-based ensemble classifier can improve CAD methods in discriminating benign from malignant lesions. |
| Mainprize et al. (76) | quantitative volumetric breast density | N/A | 0.75 | The model had an AUC of 0.75, while volumetric breast density, age-adjusted BI-RADS density, and BI-RADS density had an AUC of 0.72, 0.71, and 0.64, respectively. |
| Debelee et al. (193) | Texture features | SVM for MIAS: 100% and 96.26%.<br>SVM for BGH: 98.16% and 99.48%<br>MLP for MIAS: 75.73% and 96.65%<br>MLP for BGH: 96.26% and 97.40% | N/A | Using the MLP classifier, the fusion of features between the Gabor and CNN-extracted features acquired the maximum performance. |
| Kayode et al. (194) | 15 texture features | 91.3% and 94.4% | N/A | The automated CAD model can facilitate the diagnosis of breast cancer by radiologists and has a PPV=89.5% and an NPV=95.5% in discriminating between benign and malignant breast lesions. |
| Agnes et al. (195) | N/A | Sensitivity: 96 | 0.99 | Multiscale All Convolutional Neural Network (MA-CNN) is a strong tool for discriminating between benign and malignant breast lesions. |
| Kaur et al. (196) | N/A | N/A | 0.99 | A decision support model using Multi-SVM (MSVM) and DL has 3% improved sensitivity and 2% improved specificity compared to the Multi-Layer Perception (MLP) and J48+K-mean clustering WEKA manual approach. |
| Marino et al. (197) | Lesion geometry, the discrete Haar wavelet transform (WAV), autoregressive model, absolute gradient, run length matrix (RLM), first-order histogram (HIS), and co-occurrence matrix (COM) | N/A | N/A | The co-occurrence matrix (COM) in combination with the first-order histogram (HIS) or mutual information (MI) coefficient had an accuracy of 87.4% for differentiation of invasive from non-invasive BC. |
| Wang et al. (87) | 4,098 radiomics features | N/A | 0.93 | Radiomic features of the annular perilesional region of 3mm had the highest AUC=0.93 for distinguishing benign from malignant lesions. |
| Lin et al. (86) | 19 radiomics features | N/A | 0.94 | A radiologic nomogram combined with CES-based radiomic features and predictive factors of age and BI-RADS score had an AUC=0.94 in identifying benign from malignant breast lesions of <1cm. |
| Marino et al. (198) | N/A | N/A | N/A | DCE-MRI radiomics showed an accuracy of 90% in discriminating invasive from non-invasive BCs, while it was 92% for CEM. CESM can be used as an alternative approach if MRI is contraindicated or unavailable. |
| Fusco et al. (90) | 48 textural metrics | N/A | N/A | Considering CEM, skewness (AUC=0.71) and kurtosis (AUC=0.71) were the best predictors of malignancy. Consistently, considering features extracted from DCE-MRI, gray-level run-length matrix (AUC = 0.72), GLN (gray-level nonuniformity) (AUC = 0.72), entropy (AUC = |
| | | | | 0.70), energy (AUC = 0.72), and RANGE (AUC = 0.72) were the best predictors of malignancy. |
| Massafra et al. (83) | 464 features | 88.37% and 100% | N/A | Three different classifiers, including logistic regression, naïve Bayes, and random forest were trained on each subset of principal components. The random forest-based classifier showed the best potential to predict the malignancy of the ROIs with a sensitivity of 88.37% and a specificity of 100%. |
| Wang et al. (199) | N/A | N/A | In the external testing set: 0.947 for the radiomic-based model, 0.964 for the combined model<br>In the internal validation set: 0.921 for the radiomic model and 0.934 for the combined model | In the external test set, the radiomics models had an AUC=0.947 compared to the combined model (combining radiomics and clinical data) which had an AUC=0.964.<br>However, in the internal validation set, the combined model had an AUC=0.934, which was significantly higher than both the clinical and radiomics models (P<0.05) |
| Sakai et al. (99) | 70 features | N/A | N/A | The accuracy of the support vector machine-based classifier was 55% for benign lesions and 84% for malignant lesions. |
| Tagliafico et al. (98) | 90 percentile, skewness, and entropy | N/A | N/A | 90 percentile, skewness, and entropy were correlated with tumor size. Moreover, entropy was related to the estrogen receptor status of the BC. |
| Niu et al. (101) | N/A | Specificity of 0.946 and sensitivity of 0.970 in the training | AUC of 0.980 in the training cohort.<br>AUC of 0.985 in | The radiographic nomogram, made by incorporating menstruation status, age, and radimoic features showed a specificity of 0.946, sensitivity of 0.970, and AUC of 0.980 in the |
| | | cohort.<br>Specificity of 0.966 and sensitivity of 0.909 in the validation cohort. | the validation cohort. | training set. However, it had a specificity of 0.966, a sensitivity of 0.909, and an AUC of 0.985 in the validation cohort. |
| Peng et al. (103) | 170 features of which 26 were selected | N/A | 0.834 and 0.868 | The view-based mode and case-based mode of both 2D and 3D-domain radiomis had the highest AUC of 0.834 and 0.868, respectively |
| Alì et al. (104) | N/A | N/A | N/A | The machine learning classifier combining radiomics and BIRADS improved the PPV to 50% and reduced false-positive by a factor of 2. |
| Chao et al. (96) | N/A | N/A | GBDT had an AUC of 0.91. | GBDT model successfully differentiated benign exams from malignant ones with an AUC of 0.91. |
| Xiao et al. (105) | N/A | N/A | 2D ResNet34 (AUC=0.8264), anisotropic 3D ResNet (AUC=0.8455), and deep learning ensemble strategy (AUC=0.8837). | The ensemble CNN had improved performance and a reduction in false positive results (0.0435 increase in AUC) |
| Yang et al. (200) | N/A | N/A | N/A | The collection input-based support tensor machine (CISTM)-based classifier outperformed the kernelled support tensor machine-based classifier in classifying breast lesions. |
| Mendel et al. (100) | N/A | N/A | 0.89 | Compared to mammography, DBT did better in the classification of architectural distortions of malignant masses. |
| Luo et al. | 9 features | N/A | 0.928 | The nomogram combining radiomic features and BIRADS scoring showed a better discrimination |
| (125) | | | | power with an AUC=0.928 compared to radiomics and BIRADS scoring alone (p value=0.029 and 0.011, respectively). |
| Fleury et al. (118) | 10 features | 71.4% sensitivity, 76.9% specificity | 0.840 | The SVM had 76.9% specificity, 71.4% sensitivity, and the highest AUC of 0.840 |
| Moustafa et al. (126) | 9 grayscale and 2 color Doppler features | N/A | 0.958 | Training of the model based on both color Doppler and grayscale features improved the AUC from 0.925 to 0.958, reduced borderline diagnosis, and enhanced the diagnostic performance. |
| Du et al. (123) | 12 features | N/A | 0.986 in the training cohort and 0.977 in the validation cohort | In the training cohort and validation, the nomogram had an AUC of 0.986 and 0.977, respectively, which outperformed both clinical models and radiomics signatures. |
| Vigil et al. (151) | 354 features | N/A | N/A | The dual intended deep learning model showed an accuracy of 78.5% in the classification of sonographic breast lesions. |
| Youk et al. (106) | 730 features<br>22 grayscale US and 6 SWE features remained after feature selection | N/A | 0.929 for grayscale US and 0.992 for SWE | 21 grayscale US features and 6 SWE features were associated with malignant breast tumors. The AUC was 0.929 for grayscale US and 0.992 for SWE. |
| Sim et al. (124) | 93 features | 0.5 and 0.703 | 0.765 | In the validation set, the radiomics-based classifier had an AUC of 0.765, an accuracy of 0.703, specificity of 0.5, and sensitivity of 0.857 |
| Hassanien et al. (114) | N/A | N/A | N/A | ConvNeXt method outperformed all previously developed methods, having an accuracy of 91%. Since the quality of the US images affects the accuracy of the malignancy-predicting models, ignoring the low-quality US images improved the score pooling model's accuracy. |
| Mishra et al. (113) | N/A | N/A | 0.97 | The proposed approach showed an accuracy of 0.974 and an F1-score of 0.94 in discriminating |
| | | | | benign and malignant breast lesions. |
| Interlinghi et al. (116) | N/A | 98% and 94 % on the first and second datasets, respectively. | N/A | The proposed model reduced the rate of biopsy from 18% to 15% in benign lesions. The external validation of the model on two different datasets had a PPV of 45.9% and 98% sensitivity, compared to the 41.5% PPV of a radiologist (p<0.005) in the first dataset. Validation testing on the second dataset showed a PPV of 50.5% and sensitivity of 94%, compared to the PPV of 47.8% of a radiologist (p<0.005). |
| Michael et al. (107) | 185 features | N/A | N/A | The LightGBM classifier had an FI score of 99.80%, 99.60% recall, 100.0% precision, and 99.86% accuracy in detecting benign and malignant lesions. |
| Zhang et al. (108) | N/A | N/A | 0.96 for the test set and 0.90 for the external validation set. | This method had a sensitivity of 98.7%, an accuracy of 85.6%, and a specificity of 63.1% in the test dataset. When using the deep learning model, the same set had a sensitivity of 91.3%, an accuracy of 89.7%, and a specificity of 86.9%. |
| Zhang et al. (110) | N/A | N/A | Inception V3=0.905, VGG19= 0.847, ResNet50= 0.851, and VGG16= 0.866 | In the test group, Inception V3 had the largest AUC=0.905, VGG19 had an AUC=0.847, ResNet50 had an AUC=0.851, and VGG16 had an AUC=0.866. Compared to sonographers (AUC=0.846), Inception V3 had a higherAUC=0.913. |
| Chen et al. (122) | GLNU in the 135-degree direction, run length non-uniformity in the 135-degree direction, GLNU in the 45-degree direction, vertical gray level | 91.4% and 69.3% | 0.834 | All named texture features were higher in TNBC tumors compared to non-TNBC tumors. GLNU in the horizontal direction demonstrated the highest efficacy in differentiating TNBC from non-TNBC tumors. |
| | non-uniformity, and GLNU in the horizontal direction. | | | |
| Ternifi et al. (127) | Spatial vascularity pattern (SVP), bifurcation angle (BA), Murray’s deviation (MD), and microvessel fractal dimension (mvFD). | 89.2% sensitivity<br>93.8% specificity | 0.97 | The addition of the BIRADS score to the HDMI biomarkers produced a prediction model with 89.2% specificity, 93.8% sensitivity, and 0.97 AUC. |
| Xu et al. (121) | Histogram features | N/A | The AUC of mean and n percentiles were from 0.807 to 0.848 | Analysis of histogram-based radiomic features of US can successfully differentiate between small triple Negative breast invasive ductal carcinoma (TN-IDC) and Fibroadenoma |
| Moon et al. (120) | level-set method, morphological, conventional texture, and multiresolution gray-scale invariant | N/A | 0.9702 | The CAD using texture features extracted by based ranklet transform was practical for differentiating between benign fibroadenoma and TNBC. |
| Tang et al. (111) | N/A | N/A | 0.908 | Integration of the maximum elasticity value (Emax) with the BIRADS score resulted in the development of a model with an AUC=0.908. This improved method showed to reduce overtreatment by up to 50% and exhibited better performance than the BIRADS scoring system. |
| Romeo et al. (112) | 520 features<br>10 features were used to train the model | N/A | 0.90 in the training set and 0.82 in the test set | The radiomics-based model’s function was not better than the radiologist’s performance. And the addition of the radiomics model did not improve the radiologist's performance. |
| Niu et al. (201) | N/A | N/A | In training set: 0.975 for nomogram and 0.964 for radiomics<br>In validation set: 0.983 for nomogram and 0.978 for radiomics | Compared to DBT + mammography, DCE-MRI + DWI-MRI had a higher sensitivity and AUC, but lower sensitivity for breast cancer detection. The nomogram incorporating these features with clinical data demonstrated diagnostic advantages compared to each modality. |

**Table 1.** Summary of studies on the application of Radiomics in screening/diagnosis of breast lesions.

DWI: diffusion-weighted imaging, AUC: area under the curve, MRI: magnetic resonance imaging, N/A: not applicable, DCE: dynamic contrast-enhanced, TNBC: triple-negative breast cancer, NPV: negative predictive value, PPV: positive predictive value, CAD: computer-aided diagnosis, CNN: convolutional neural network, BIRADS: breast imaging-reporting and data system, CESM: Contrast-enhanced spectral mammography, DBT: Digital breast tomosynthesis, GBDT: gradient boosting decision tree, SVM: support vector machine, US: ultrasonography, SWE: shear-wave elastography, GLNU: Gray level non-uniformity

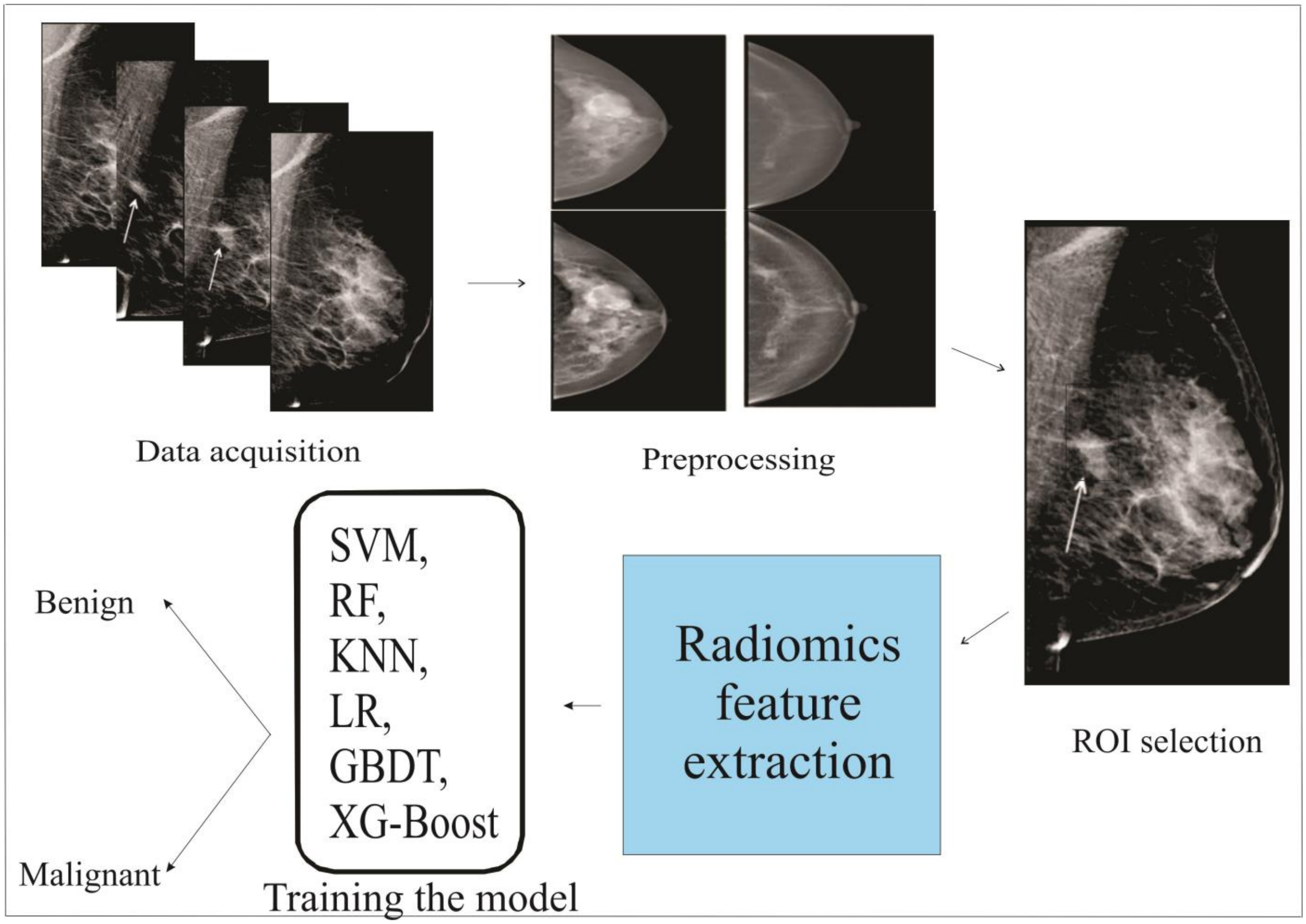

**Figure 1.** The radiomics-based machine learning workflow for diagnosis of breats cancer.

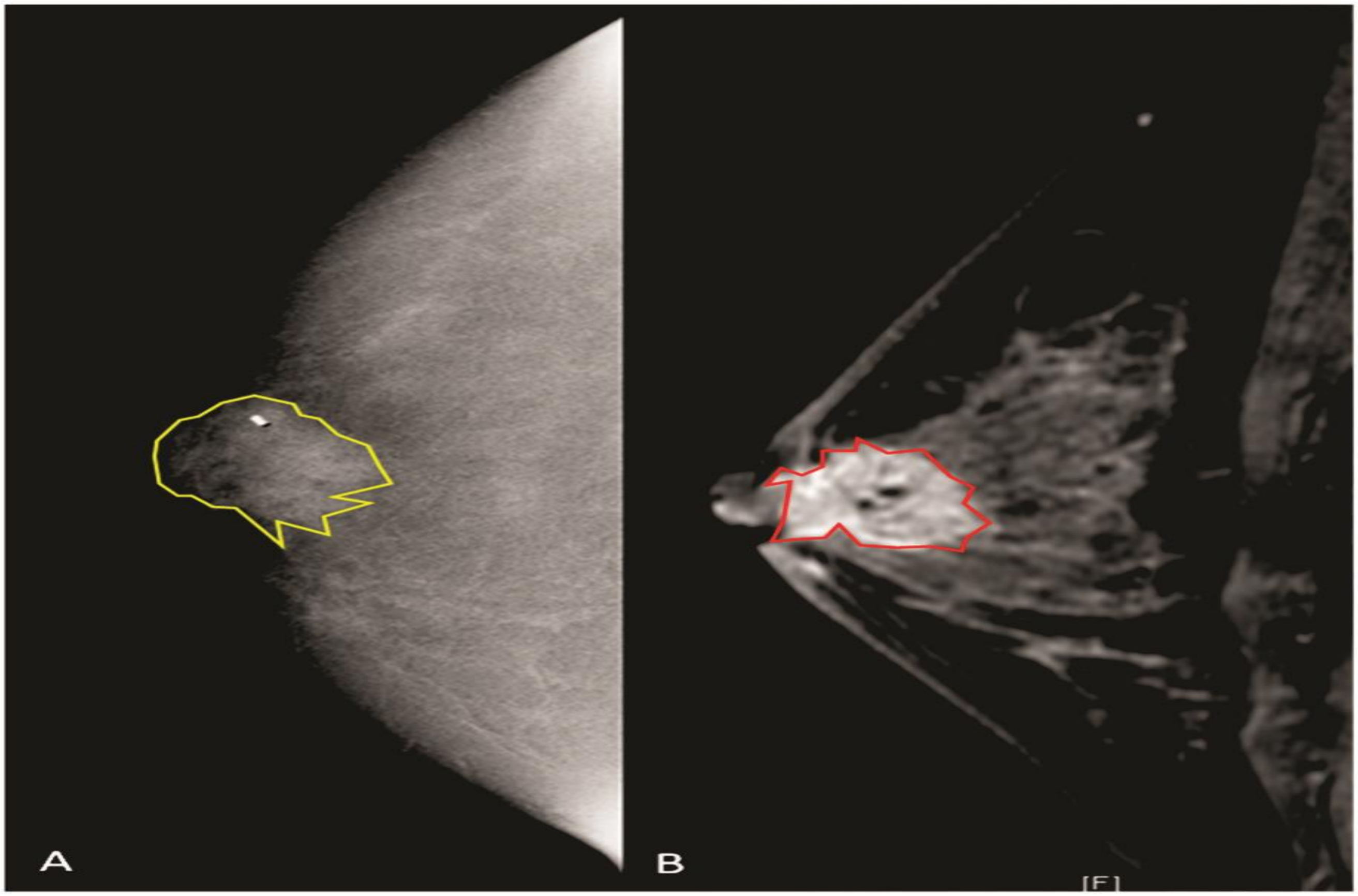

**Figure 2.** Biopsy-proven invasive ductal carcinoma (G3, HR positive, HER2 negative) in a 42-year old women. (A) Contrast-enhanced mammography shows, on the right, a 22 mm rounded area of enhancement. The lesion was manually segmented (2D yellow region of interest), and the clip marker had to be included in the segmented area. (B) Sagittal fat-saturated post-contrast enhanced T1-weighted image. The breast is heterogeneously dense with moderate background enhancement. In the retroareolar right breast, there is a 22 mm spiculated mass containing a localizing clip from biopsy. The mass is inseparable from the nipple, which appears slightly retracted, and there is associated skin thickening. The lesion was manually segmented (2D red region of interest) (85).

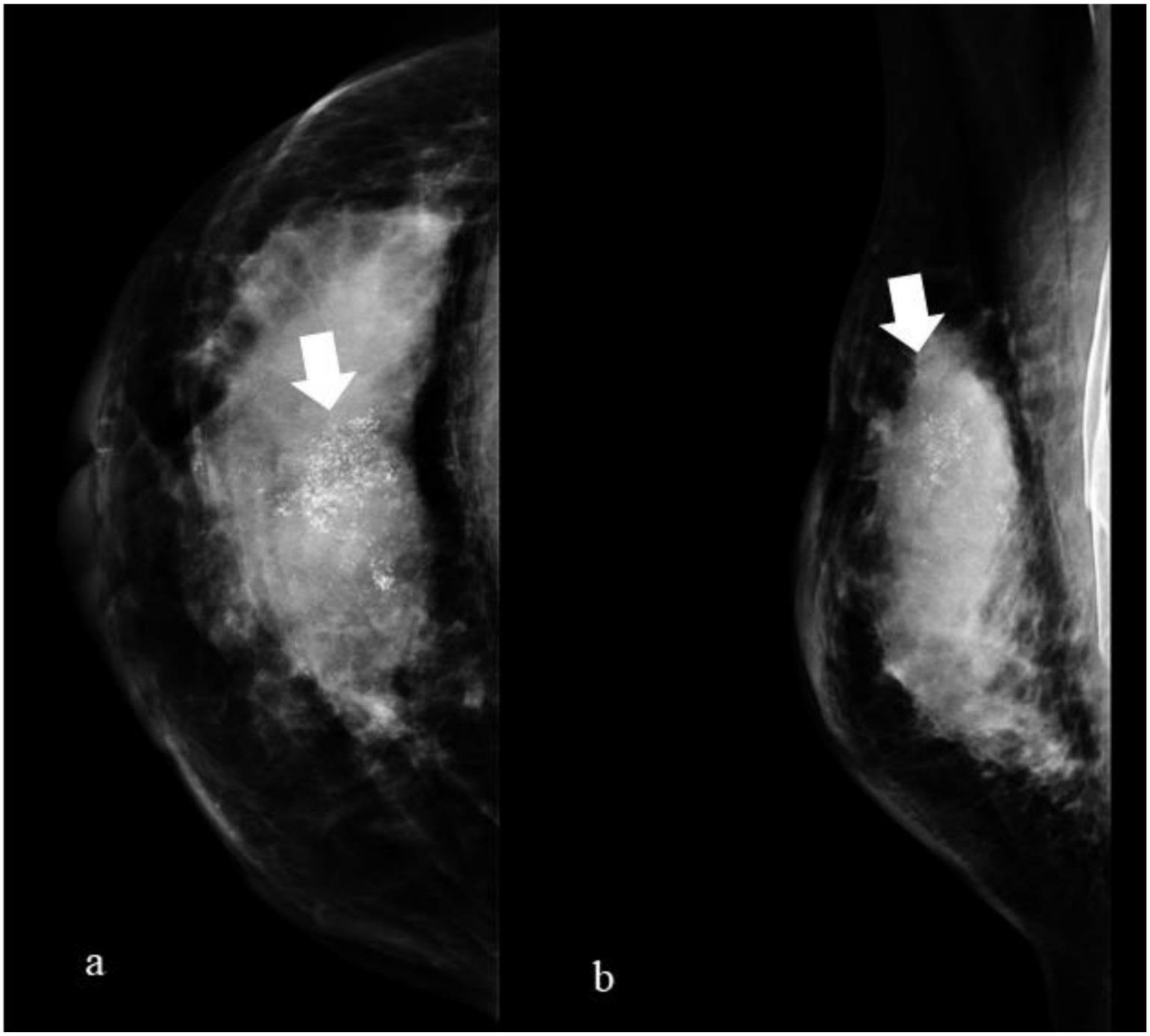

**Figure 3.** A 45-year-old woman with infiltrating ductal carcinoma of the right breast. (a) a craniocaudal and mediolateral oblique projection (b) radiogram of the right breast with the presence of polymorphic microcalcifications extended to the upper sectors (arrow) (202).

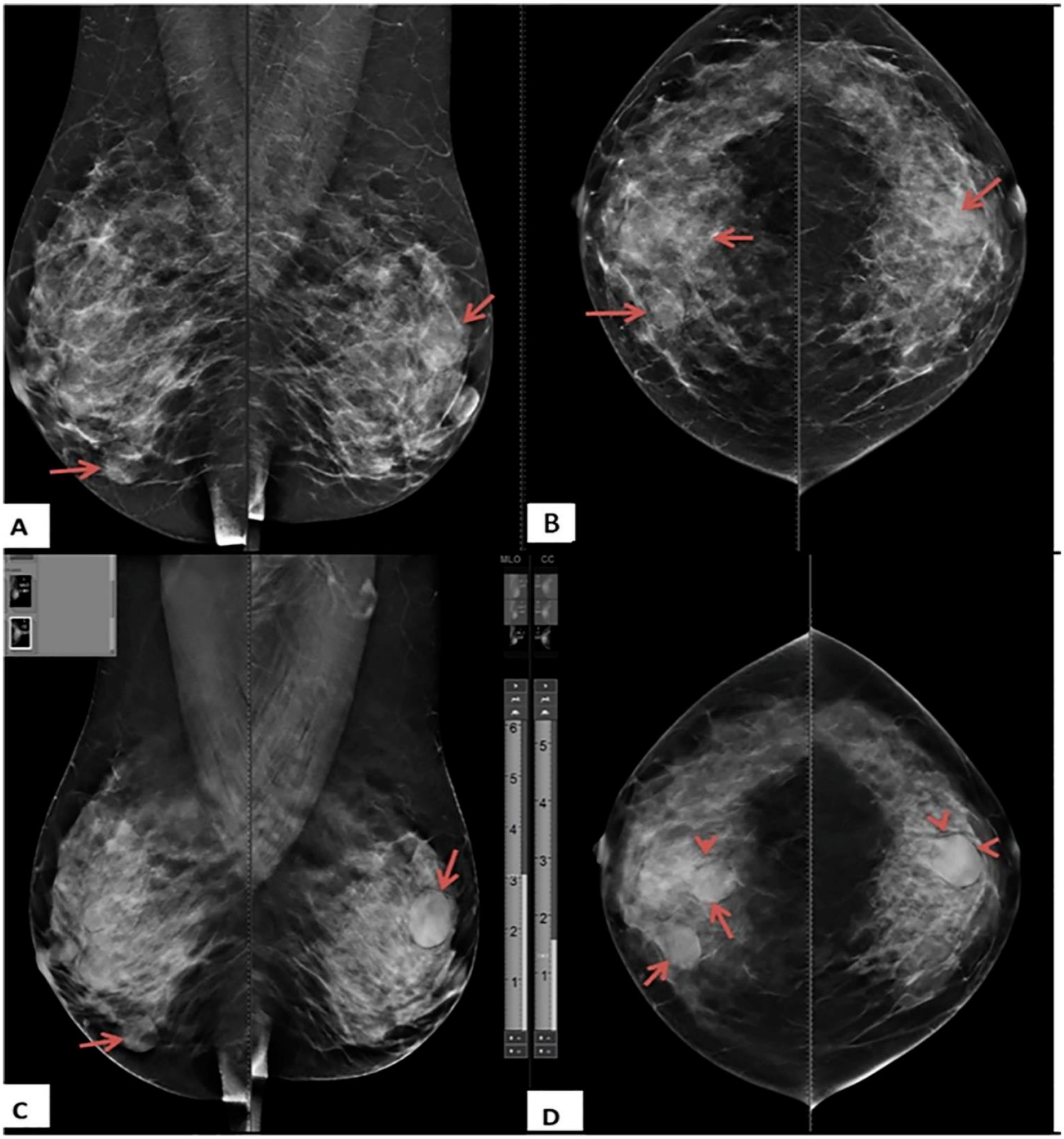

**Figure 4.** Digital Mammography and Digital Breast Tomosynthesis imaging in a 40-year-old female presented with bilateral breast lumps. (A) Mediolateral oblique and (B) Craniocaudal Digital Mammography images of both breasts show heterogeneous dense breasts (ACR C) with left breast

retroareolar, right breast retroareolar, and right breast lower inner quadrant dense lesions with obscured margins (arrows). No microcalcifications or speculated masses. (C) Mediolateral oblique and (D) Craniocaudal Digital Breast Tomosynthesis images show more margin characterizations of the lesions, which are medium-dense, well-defined rounded lesions with smooth margins and a characteristic halo sign (arrowheads). The lesions were classified as BI-RADS 3 according to Digital Mammography and BI-RADS 2 according to Digital Breast Tomosynthesis. Histopathological examination revealed simple cysts (203).

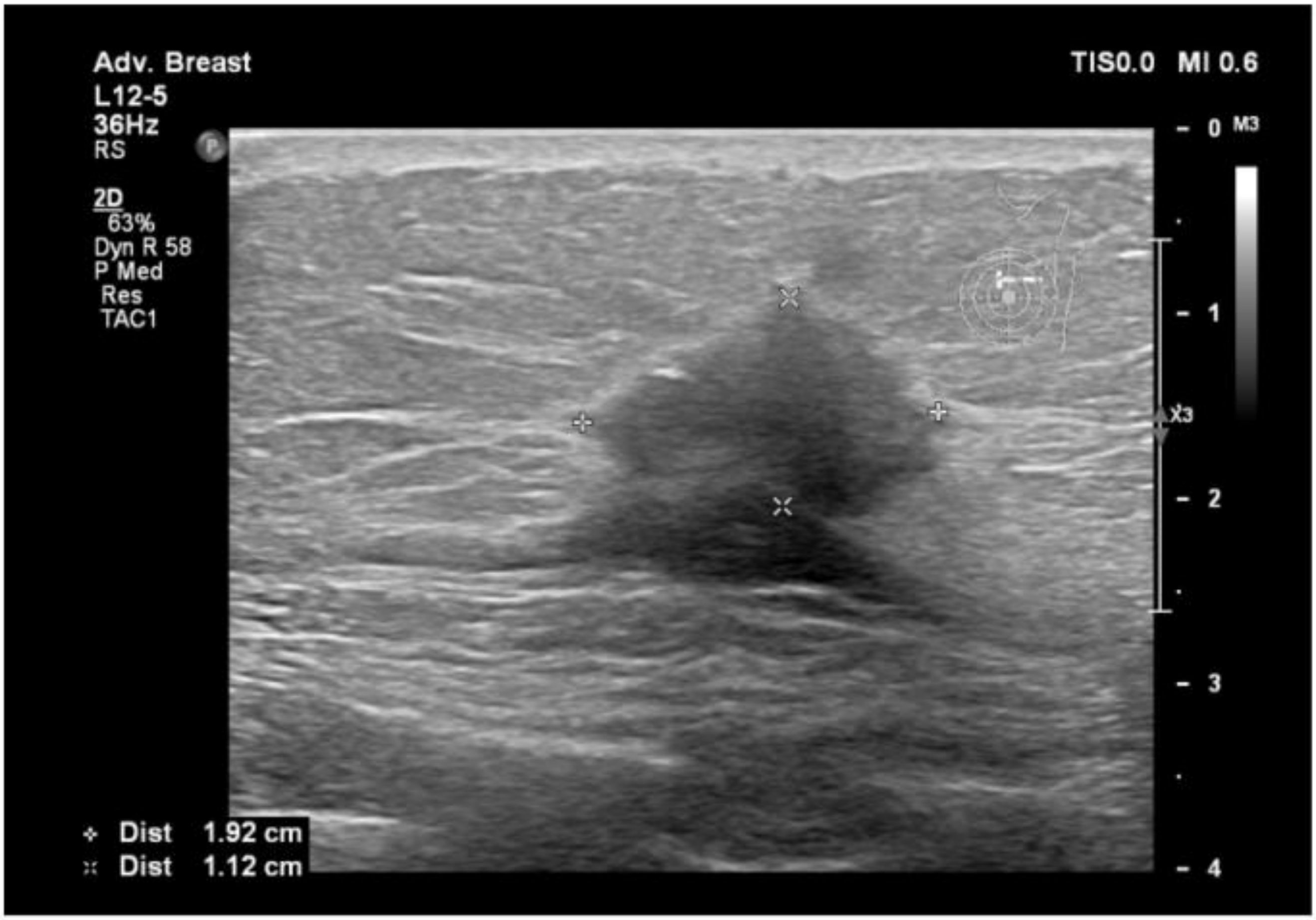

**Figure 5.** Breast lesion detected using ultrasound, shows a hypoechogenic solid nodule, with spiculations and ill-defined margins (204).